# Drift-insensitive distributed calibration of probe microscope scanner in nanometer range: Virtual mode

Rostislav V. Lapshin[1, 2]

[1]*Solid Nanotechnology Laboratory, Institute of Physical Problems, Zelenograd, Moscow, 124460, Russian Federation*

[2]*Moscow Institute of Electronic Technology, Zelenograd, Moscow, 124498, Russian Federation*

E-mail: rlapshin@gmail.com

A method of distributed calibration of a probe microscope scanner is suggested which main idea consists in a search for a net of local calibration coefficients (LCCs) in the process of automatic measurement of a standard surface, whereby each point of the movement space of the scanner can be characterized by a unique set of scale factors. Feature-oriented scanning (FOS) methodology is used as a basis for implementation of the distributed calibration permitting to exclude *in situ* the negative influence of thermal drift, creep and hysteresis on the obtained results. Possessing the calibration database enables correcting in one procedure all the spatial distortions caused by nonlinearity, nonorthogonality and spurious crosstalk couplings of the microscope scanner piezomanipulators. To provide high precision of spatial measurements in nanometer range, the calibration is carried out using natural standards – constants of crystal lattice. One of the useful modes of the developed calibration method is a virtual mode. In the virtual mode, instead of measurement of a real surface of the standard, the calibration program makes a surface image "measurement" of the standard, which was obtained earlier using conventional raster scanning. The application of the virtual mode permits simulation of the calibration process and detail analysis of raster distortions occurring in both conventional and counter surface scanning. Moreover, the mode allows to estimate the thermal drift and the creep velocities acting while surface scanning.

## 1. Introduction

By using several techniques embedded into the feature-oriented scanning (FOS) methodology [1, 2], a new distributed approach to calibration of the probe microscope scanner is suggested [3, 4]. The essence of the developed approach is that instead of characterizing the whole movement space of a probe microscope scanner by three calibration coefficient $K_x$, $K_y$, $K_z$ [5], each point $(x, y, z)$ of this space is characterized by its own unique triplet of local calibration coefficients (LCCs) $K_x(x, y, z)$, $K_y(x, y, z)$, $K_z(x, y, z)$ [4]. As a result, it is possible to correct all spatial distortions caused by nonlinearity, nonorthogonality and spurious crosstalk couplings of the microscope scanner X, Y, Z piezomanipulators. In the real mode [6], application of the FOS approach [1, 2] and the methods of counter-scanning [7] permits eliminating *in situ* the negative influence of thermal drift, creep, and hysteresis on the distributed calibration results.

A reference surface used for calibration should consist of elements, called hereinafter features, such that the distances between them or their sizes are known with a high precision. The corrected coordinate of a point on the distorted image of an unknown surface is obtained by summing up the LCCs related to the points of the movement

**Drift-insensitive distributed calibration of probe microscope scanner**

trajectory of the scanner [4].

Virtual distributed calibration is such a calibration that the physical surface of a standard is substituted with a topography image under correction. Of course, the thermal drift and creep of the microscope can not possibly have been excluded during such sort of calibration, which is reflected in the obtained LCC and obliquity angle distributions. After the virtual calibration database (CDB) of the image under correction has been created, the nonlinear correction of the image is carried out. It does not matter for this mode which exact factors have distorted the scan under correction, e. g., thermal drift, creep, or static piezoscanner nonlinearities acting together or separately.

The virtual mode is intended for simulating the process of calibration and validating the analytical solutions found in Ref. 4. The virtual mode of distributed calibration allows thorough analysis of probe microscope scanner operation. In particular, the mode permits determination of the values and the character of raster distortions for both regular and counter types of scanning. Moreover, the virtual mode can be used to estimate the thermal drift and creep velocities, for moiré detection, and for automatic characterization of crystal surfaces.

## 2. Measurement conditions

The atomic topography of the basal plane (0001) of highly oriented pyrolytic graphite (HOPG) monocrystal was used as a standard surface. The measurements were carried out at the scanning probe microscope (SPM) Solver™ P4 (NT-MDT Co., Russia) by method of scanning tunneling microscopy (STM) in the air at room temperature. In order to minimize thermal deformation of the sample, a graphite crystal of small dimensions 2×4×0.3 mm was used. Three adjacent carbon atoms (or interstices) forming an equilateral triangle ABC (see Fig. 3(a) in Ref. 4) were selected as a local calibration structure (LCS). According to the neutron diffraction method, the HOPG lattice constant $a$ (i. e., side length of the ABC triangle) makes 2.464±0.002 Å [8].

As the tip, a mechanically cut ⌀0.3 mm NiCr wire was used. To protect the microscope against seismic oscillations, a passive vibration isolation system was employed. Moreover, the microscope was housed under a thermoinsulation hood, which also served as an absorber of external acoustic disturbances. The typical noise level of the tunneling current in the course of the measurements made about 20 pA (peak-to-peak).

During the raster scanning, the probe movement velocity at the retrace sweep was set the same as at the forward trace. Immediately before the raster scanning begins, scanner "training" was carried out [7]. The scanner training is a repeated movement along the first line, which allows to decrease creep at the beginning of the scan [9]. While training, the actual scanning velocity was also determined.

Some of the values in the sections presented below are intentionally given with a redundant number of significant digits. That will permit to compare them with the similar values obtained under different measurement conditions or in different measurement modes.

## 3. Almost linear raster distortions

*3.1. Analysis, correction, comparison of errors of direct and counter images*

STM-scans of HOPG surface distorted by drift are shown in Fig. 1. A regular (direct) surface scan and a scan counter to it [7] are given in Fig. 1(a) and Fig. 1(b), respectively. Virtual FOS [1] of the presented images allows determining a mean atomic lattice spacing as 2.734±0.25 Å and 2.199±0.20 Å, whence it is easy to estimate a relative measurement error as 11.0% and 10.8%, respectively. Mean spacings determined by interstices made 2.731±0.25 Å and 2.197±0.20 Å, relative measurement error is 10.8% and 10.9%, respectively.

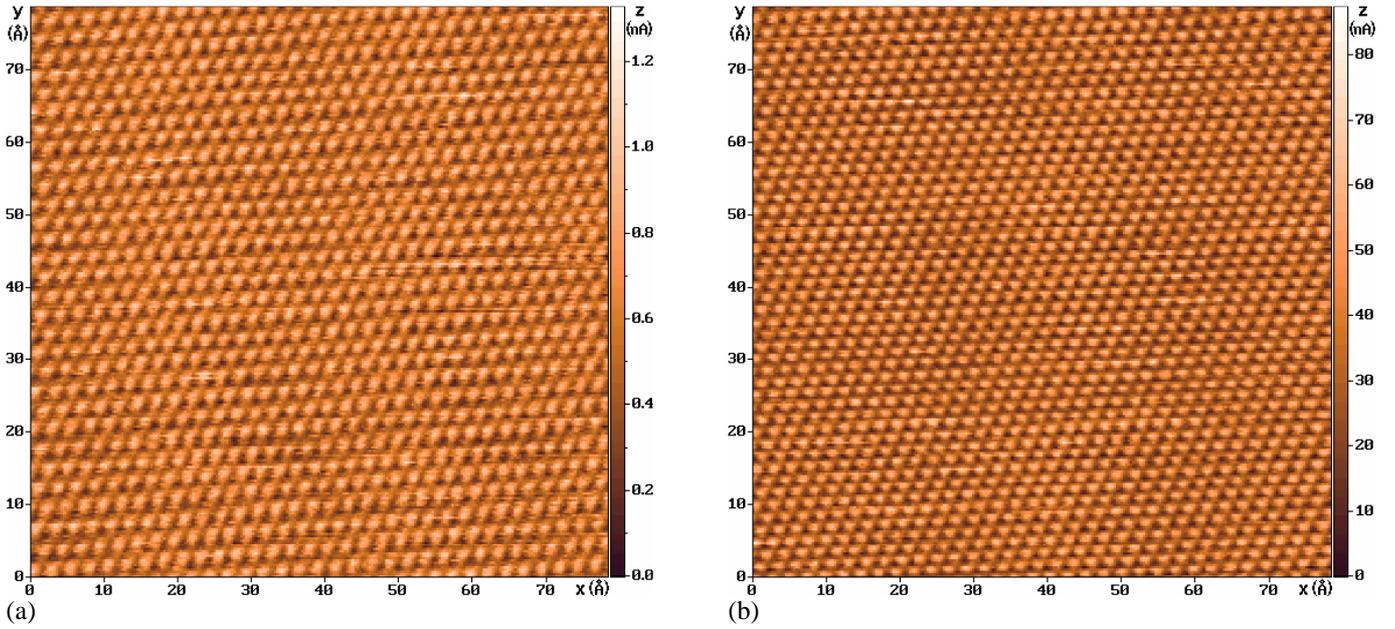

Fig. 1. Drift-distorted scans of atomic surface of pyrolytic graphite (a) direct image, (b) counter image. Measurement mode: STM, constant-height, $U_{tun}$=85 mV, $I_{tun}$=750 pA. Number of points in the raster: 256×256. Scanning step size: $\Delta_x$=0.306 Å, $\Delta_y$=0.307 Å. Number of averagings at the raster point is 15. Scanning velocity $v_x$=$v_y$=223.1 Å/s (is determined while training). CSI scanning time $T_{CSI}$=6 min. Mean constant of atomic lattice equals to (a) 2.7 Å, (b) 2.2 Å which corresponds to the relative measurement error of (a) 11%, (b) 11%.

In Fig. 2, shown are: partitioning of the graphite scan by an integer-valued net having square cells; "probe travellings" during the virtual calibration (compare with the trajectory of the real mode in Ref. 6); and LCS positions for which LCCs and local obliquity angles were determined. The images in Fig. 2 correspond to the case when carbon atoms are used as features; using interstices would give a similar result. In order to decrease the influence of edge effects [9, 10], features located along the image edges were excluded from consideration by setting corresponding area of distributed calibration.

Apertures of 37×37 points and segments of 25×25 points were used while calibrating by the direct image; 31×31 and 21×21 – while calibrating by the counter one. 1296 apertures were "scanned" on the direct image and 1849 apertures – on the counter one. All LCSs found in the aperture were used during the calibration. The calibration program revealed no defective LCSs.

The number of LCSs found, the mean values of lateral LCCs and the obliquity angle of the virtual CDB are given in Table 1. It is seen well from the table presented that calibration by LCSs consisted of carbon atoms and calibration by LCSs consisted of interstices result in very close values. Therefore, at least with the accuracy of tenth of a percent, the interstices may be used as features for scanner calibration along with the carbon atoms and the CDBs obtained by both types of features may be combined into a single CDB.

It should be noted that despite of a noticeable difference in mean coefficients $<\overline{K}_y>$ of the direct and the counter scans from the initial lumped calibration coefficient $\Delta_y$=0.307 Å [7], we can speak about approximately equal deviation of these coefficients from coefficient $\Delta_y$, i. e., the relation $\Delta_y/<\overline{K}_y>$ for the direct scan is approximately equal to the relation $<\overline{K}_y>/\Delta_y$ for the counter scan.

Second order regression surfaces $\overline{K}_x^r$, $\overline{K}_y^r$, $\alpha^r$ built by the virtual CDB of the direct image are shown in Figs. 3(a)-(c) (CDB obtained by LCSs consisting of carbon atoms and CDB obtained by LCSs consisting of inter-

**Drift-insensitive distributed calibration of probe microscope scanner**

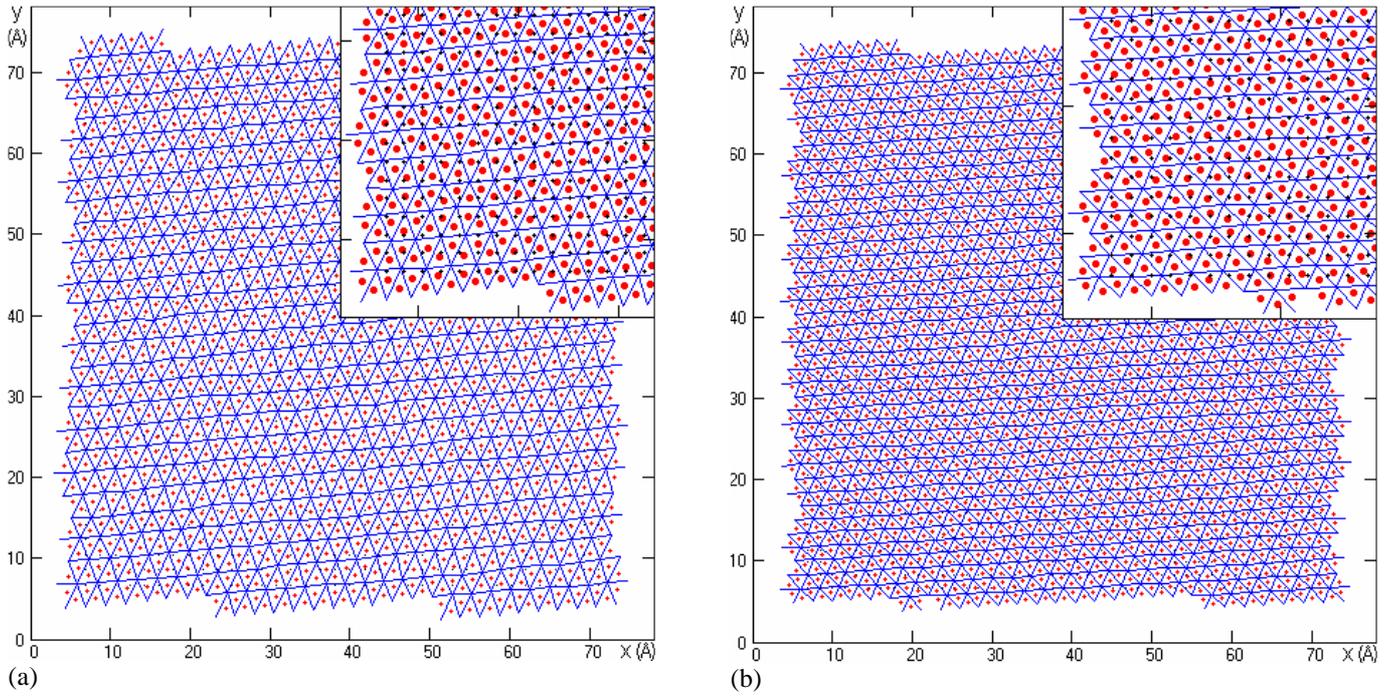

Fig. 2. LCS positions (designated as "●") for which LCCs and local obliquity angles were obtained during the virtual distributed calibration by (a) direct, (b) counter image. Trajectory of "probe traveling" is imaged with a solid line. Initial net (a) 36×36, (b) 43×43 nodes, step (a) 6, (b) 5 positions. Net nodes are designated with a "+" symbol. Carbon atoms are used as features. Number of processed apertures (a) 1296, (b) 1849. Number of found LCSs (a) 1503, (b) 2274. The inset shows a zoomed area located in the lower left corner.

stices are integrated into a single CDB). As the scans under analysis have small sizes, such distortions as spurious couplings between manipulators and static piezoscanner nonlinearity [6] would be insignificant on the regression surfaces.

Indeed, regression surface $\overline{K}_x^r$ (see Fig. 3(a)) is parallel to axis $y$ indicating that the scanner movement along the "slow" scan direction does not impact the scanner movement along the "fast" scan direction (LCCs $\overline{K}_x$ are changing equally in all raster lines). Regression surface $\overline{K}_y^r$ (see Fig. 3(b)) is parallel to axis $x$ indicating that the scanner movement along the "fast" scan direction does not impact the scanner movement along the "slow" scan direction (LCCs $\overline{K}_y$ are changing equally in all raster columns). Almost horizontal position of regression surface $\alpha^r$, which is parallel to $x$ axis and slightly tilted to $y$ axis (see Fig. 3(c)), reveals a weak dependence of the obliquity angle on the scanner movements in the lateral plane. The similar situation is observed for the counter image (see Figs. 3(d)-(f)). Thus, the creeps excited in the X segments and Y segments of the piezotube scanner have practically no couplings with each other, at least in the movement scales being considered.

Table 1. Mean values of lateral LCCs and obliquity angles of virtual CDBs. The first value is a result of calibration by carbon atoms, the second one – by interstices of crystal lattice.

| HOPG scan | Number of LCSs in CDB | $<\overline{K}_x>$ (Å) | $<\overline{K}_y>$ (Å) | $<\alpha>$ (deg) |
|---|---|---|---|---|
| Direct, Fig. 1(a) | 1503, 1496 | 0.3075, 0.3084 | 0.2537, 0.2538 | -9.40, -9.38 |
| Counter, Fig. 1(b) | 2274, 2261 | 0.3077, 0.3078 | 0.3900, 0.3921 | 7.08, 7.02 |
| Direct, Fig. 6(a) | 1233, 1188 | 0.3072, 0.3070 | 0.2379, 0.2387 | -25.40, -24.99 |
| Counter, Fig. 6(b) | 2248, 2174 | 0.3078, 0.3079 | 0.4074, 0.4063 | 8.99, 8.92 |

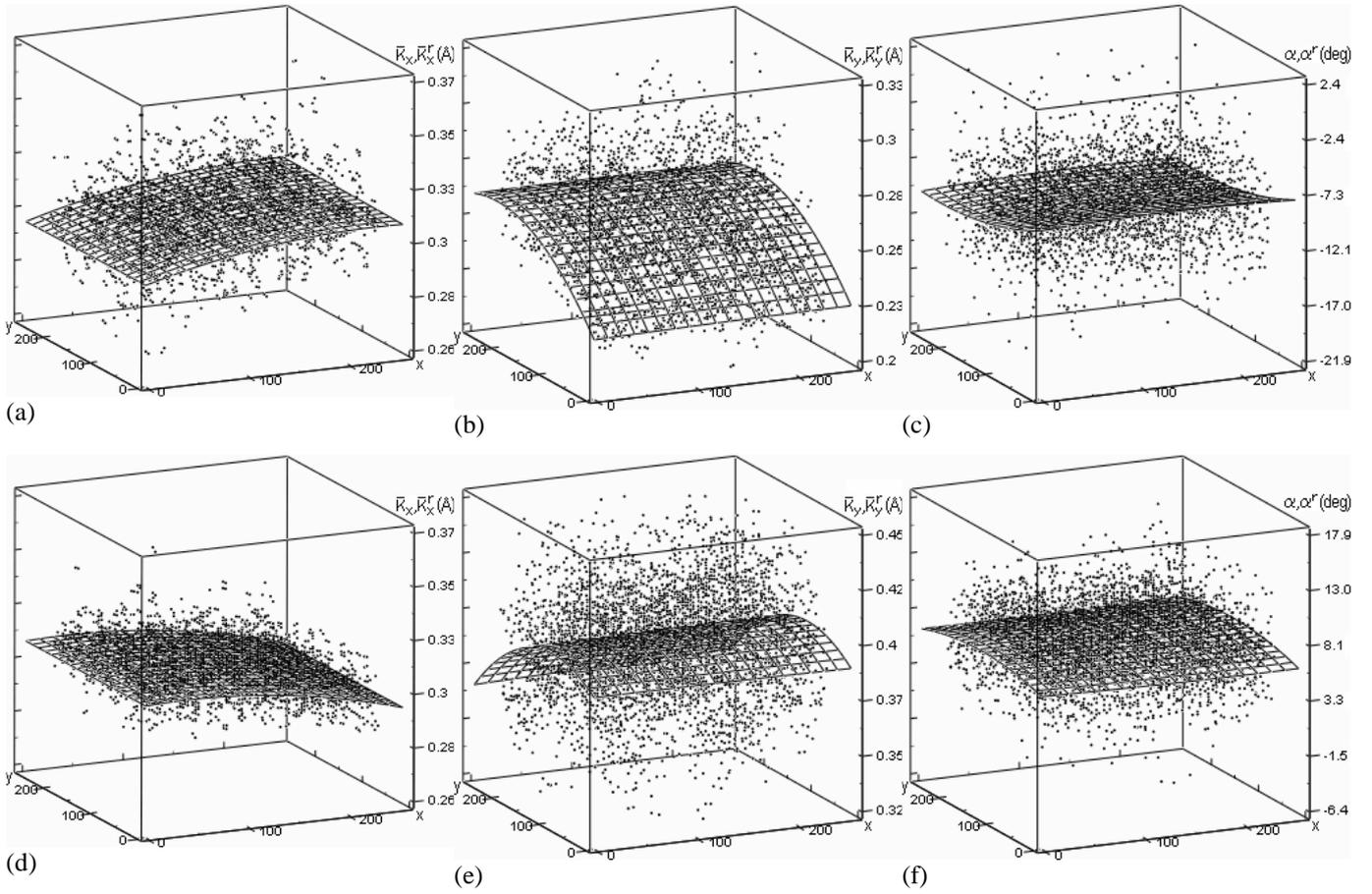

Fig. 3. Regression surfaces of 2nd order drawn through (a), (d) LCCs $\overline{K}_x$, (b), (e) LCCs $\overline{K}_y$, (c), (f) local obliquity angles $\alpha$ of the direct (upper row) and the counter (lower row) images. Joint CDB is used which includes LCSs of atoms and LCSs of interstices. Vertical scales of the corresponding figures in upper and lower rows are the same.

The estimate of root-mean-square deviations of LCCs and the obliquity angle shown in Fig. 3 is carried out by formulae (14) given in Ref. 4 and the obtained values are presented in Table 2. In particular, it is clear from the table that, regardless of whether it is a direct scan or counter, the dispersion of LCCs $\overline{K}_x$ is less than the dispersion of LCCs $\overline{K}_y$ [11]. Such ratio of dispersions can be easily explained – any change in drift velocity during the raster scanning has a greater impact in the slow scan direction than in the fast one [11, 12].

Although, comparing the mean values of LCCs $\overline{K}_x$ of the direct and the counter images (see Table 1) indicates

Table 2. Standard deviations of LCCs and obliquity angles of the virtual CDBs. The first value corresponds to the regression surface of the 1st order, the second – 2nd order.

| HOPG scan | $\sigma_{\overline{K}_x}$ (Å) | $\sigma_{\overline{K}_y}$ (Å) | $\sigma_\alpha$ (deg) |
|---|---|---|---|
| Direct, Fig. 1(a) | 0.0123, 0.0123 | 0.0158, 0.0153 | 3.18, 3.18 |
| Counter, Fig. 1(b) | 0.0075, 0.0074 | 0.0251, 0.0247 | 2.48, 2.46 |
| Direct, Fig. 6(a) | 0.0101, 0.0101 | 0.0108, 0.0107 | 4.06, 3.12 |
| Counter, Fig. 6(b) | 0.0068, 0.0067 | 0.0209, 0.0208 | 2.27, 2.26 |

**Drift-insensitive distributed calibration of probe microscope scanner**

Table 3. Estimate of nonlinear distortions of the counter-scanned images expressed in terms of the maximal deviation of a regression surface from the horizontal plane. The first value corresponds to the 1st order regression surface, the second value – 2nd order regression surface. The estimate percentage is given in parentheses.

| HOPG scan | $\Delta^{max}_{\overline{K}^r_x}$ (Å) | $\Delta^{max}_{\overline{K}^r_y}$ (Å) | $\Delta^{max}_{\alpha^r}$ (deg) |
|---|---|---|---|
| Direct, Fig. 1(a) | 0.014 (100%), 0.015 (100%) | 0.034 (100%), 0.039 (100%) | 2.16 (100%), 2.42 (100%) |
| Counter, Fig. 1(b) | 0.013 (93%), 0.014 (93%) | 0.022 (65%), 0.034 (87%) | 0.23 (11%), 0.93 (38%) |
| Direct, Fig. 6(a) | 0.015 (100%), 0.017 (100%) | 0.040 (100%), 0.041 (100%) | 26.87 (100%), 32.12 (100%) |
| Counter, Fig. 6(b) | 0.013 (87%), 0.014 (82%) | 0.032 (80%), 0.034 (83%) | 2.36 (9%), 2.85 (9%) |

that the creep equally distorts these images in raster line, comparing the $\overline{K}_x$ distributions in Fig. 3(a) and Fig. 3(d) (see also Table 2) indicates that the variance of these coefficients in the counter image is notably less. This fact, in its turn, means that LCS in the counter image and, therefore, the image as a whole are less subjected to disturbing factors, mainly, such as creep.

As to the variance of $\overline{K}_y$ distribution, contrariwise, it is somewhat larger in the counter image than in the direct one. Most likely this is due to the unequal representation of the features in those images along the slow scan direction. The fact is that the image of atom/interstice in the counter scan has fewer points along the slow scan direction (see Figs. 1, 2 and 2nd column of Table 1).

Nonorthogonalities of the scanner shown in Table 1 are mainly connected with the drift-caused raster distortion (shift of lines relative to each other). The actual nonorthogonality of the scanner is tens of times less ($\alpha$=0.4° [7], see also Ref. 6). Thus, the value of obliquity angle can distinctly point out the presence of a drift. The table shows that the absolute value of mean obliquity angle of the counter image is less than the mean obliquity angle of the direct image. This means that the counter image is less distorted. Moreover, like with the distribution of LCCs $\overline{K}_x$, the variance of the distribution of the obliquity angles in the counter image is also less (cp. Fig. 3(c) with Fig. 3(f), see Table 2).

The surface $\overline{K}^r_x$ enters into transformations (5) given in Ref. 4 linearly, therefore, a slight change of this surface would lead to a slight nonlinear contribution to the raster distortion along *x*. The surface $\overline{K}^r_y$ also enters into transformations (5) linearly but it has a more distinct change, therefore, its influence along *x* and *y* directions is nonlinear to a greater extent. The observed tilt of the regression surfaces is linked with creep (see Sec. 3.2) [7]. The surface $\alpha^r$ enters into transformations (5) nonlinearly. Nevertheless, considering that $\alpha^r$ is practically constant all over the image field and is quite small ($\sin(\alpha^r) \approx \alpha^r$, $\cos(\alpha^r) \approx 1$), its contribution to the nonlinear distortion along *x* and *y* is insignificant.

Visually, the regression surfaces of the counter image (see Figs. 3(d)-(f)) look like the surfaces of the direct

R. V. Lapshin

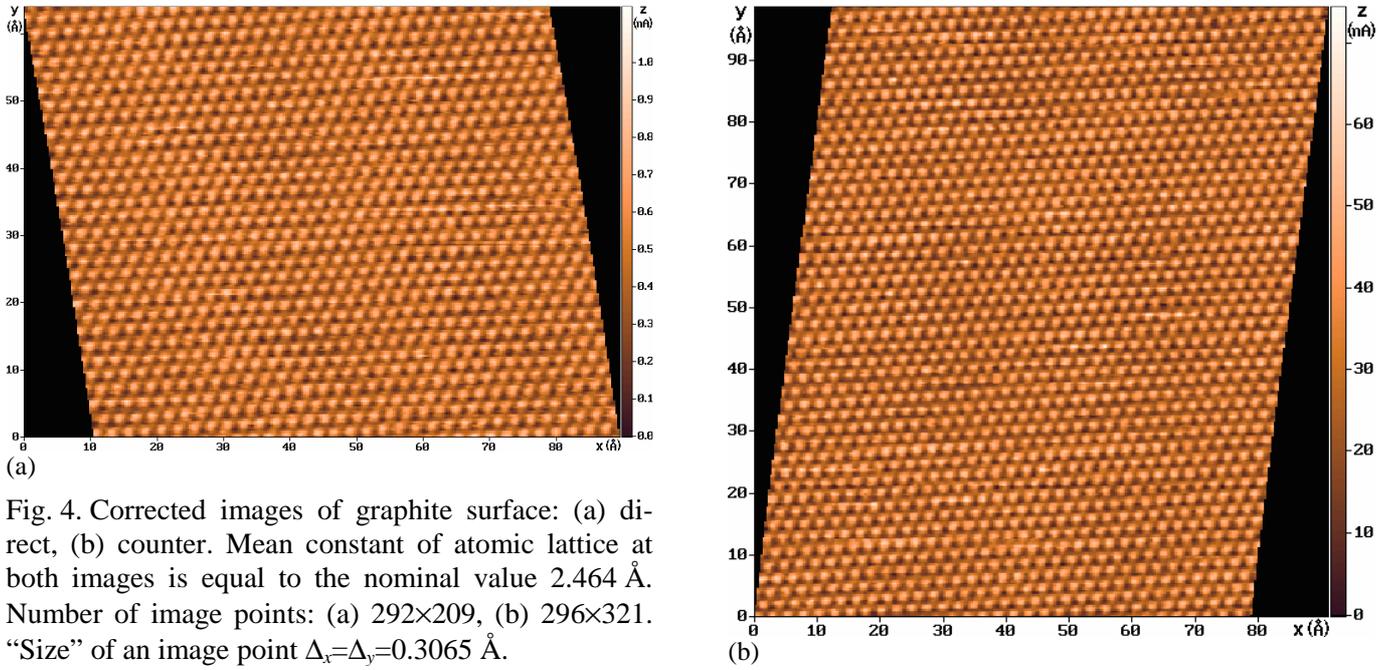

Fig. 4. Corrected images of graphite surface: (a) direct, (b) counter. Mean constant of atomic lattice at both images is equal to the nominal value 2.464 Å. Number of image points: (a) 292×209, (b) 296×321. "Size" of an image point $\Delta_x=\Delta_y=0.3065$ Å.

image but they are tilted to the opposite side. In order to quantify distinctions between direct and counter images, part of which has been discussed above (see Tables 1, 2), the maximum differences (see Table 3) were calculated for regression surfaces shown in Fig. 3 by formulae (13) given in Ref. 4. According to the table data, the counter scan is distorted less than the direct one (the estimate for distortion of the direct scan is taken for 100%).

The corrected counter-scanned images (CSIs) of graphite surface are shown in Fig. 4. Correction of the direct image is carried out by formulae (8) given in Ref. 4 with the use of the 1st order regression surfaces $\overline{K}_x^r$, $\overline{K}_y^r$, $\alpha^r$. As regards the 2nd order of the regression surfaces in Fig. 3, it was involved mainly just to point out the fact that these surfaces are close to planes. Correction of the counter image is carried out by formulae (8), where the 2nd order regression surfaces (11) given in Ref. 4 are used instead of $\overline{K}_y^r(x,y)\sin[\alpha^r(x,y)]$, $\overline{K}_y^r(x,y)\cos[\alpha^r(x,y)]$; the order of surface $\overline{K}_x^r$ is equal to 1. With only the mentioned types and powers of the regression polynomials, the nonlinear correction provides the least residual errors. The estimate of the residual errors is carried out by a mean value of the lattice constant, which is determined during virtual FOS [1] of the corrected image (see Table 4).

It should be noted that to obtain such small deviations of the mean lattice constant from its nominal value, it is necessary to use rather large samples (see column 2 containing the number of involved atoms), i. e., the scans should be of size no less than the ones shown in Fig. 1. Since the counter image contains approximately 1.5 times as many features as the direct one, in order for the residual errors of these images to be compared correctly, the mean lattice constant of the counter image should be estimated with using the same number of carbon atoms as in the direct image. However, considering that the carbon atoms are represented in the counter image with less number of points and therefore their positions are determined less accurately, all available atoms in the counter image were used to estimate the mean lattice constant.

Beside the equality of the mean lattice constant $a$ in the direct and counter images to the nominal value, other evidences of the validity of the implemented correction are (see Fig. 4): the same orientation of the crystallographic directions on the direct and the counter images and the absence of curvature of these directions. When necessary, the corrected CSIs in Fig. 4 can be superposed into a single image by using a coincidence point [7]. While super-

**Drift-insensitive distributed calibration of probe microscope scanner**

Table 4. Mean values of lattice constant in the corrected CSIs of pyrographite (in parentheses, the relative correction errors are given). Mean values of LCCs (normalized to the lateral "size" of point of the corrected image) and obliquity angles of CDBs obtained during the virtual calibration by the corrected CSIs.

| HOPG scan | Number of C atoms | Mean lattice constant (Å) | Number of LCSs in CDB | $<\overline{K}_x>\Delta_x^{-1}$ | $<\overline{K}_y>\Delta_y^{-1}$ | $<\alpha>$ (deg) |
|---|---|---|---|---|---|---|
| Direct, Fig. 4(a) | 588 | 2.4641±0.12 (0.002%) | 1454, 1438 | 1.0005 | 1.0006 | 0.15 |
| Counter, Fig. 4(b) | 846 | 2.4640±0.09 (0.001%) | 2255, 2215 | 1.0009 | 1.0009 | -0.05 |
| Direct, Fig. 9 | 331 | 2.4640±0.07 (0.001%) | 1007, 1032 | 1.0022 | 1.0037 | 0.40 |
| Counter | 845 | 2.4653±0.11 (0.052%) | 2330, 2314 | 1.0010 | 1.0004 | -0.47 |

posed, the topography is averaged within the overlap region of these images resulting in additional noise level reduction in the obtained image.

The validity of the suggested nonlinear correction was additionally confirmed by performing the virtual distributed calibrations by the corrected images of Fig. 4 (self-consistency test). The evidences of the validity (see Table 4) are a near-unit relation of the mean value of lateral LCC to the lateral "size" $\Delta_x=\Delta_y=0.3065$ Å of a point of the corrected image, a near-zero mean value of the obliquity angle, and the regression surfaces approaching horizontal planes [4].

*3.2. Determination of thermal drift and creep velocities*

By subtracting from the $x$, $y$ sizes of the corrected image (see Fig. 4) the $x$, $y$ sizes of the initial scan (see Fig. 1), respectively, and dividing it by the known scanning time of the image $T=T_{CSI}/2$, the $x$, $y$ components of mean velocity of total drift (thermal drift + creep) $\overline{v}_x$, $\overline{v}_y$ are to be found (see Table 5). Numerically, the obtained drift velocities correspond to the drift velocities observed during the real calibration after scanner calming down in the current net node (waiting state + a sequence of attachments, see Ref. 6). Since the obtained drift velocities have the same direction regardless of whether the scan is direct or counter, the thermal component makes the main contribution to the drift [7].

Also, the obtained drift velocities approximately correspond to the drift velocities presented earlier in Ref. 7. Since doubling the scan sizes in Fig. 1 relative to the scan sizes in Figure 3 of Ref. 7 did not change the drift velocity, the scanner experiencing heating due to internal losses during scanning [13] is not the main source of thermal drift in the given microscope while scanning with atomic resolution.

Suppose that the thermal drift velocity is not changing while scanning [1, 7]. The fact (see Fig. 3(b), Fig. 5(b) and equations (8) in Ref. 4) that the values belonging to the regression surface $\overline{K}_y^r$ of the direct image are less than the specified lumped coefficient $\Delta_y=0.307$ Å [14] means that during the scanning the actual microscope step while moving from line to line was less than the specified one. Step decrease occurs when thermal drift $\overline{v}_y^t$ is directed along the movement. Step decrease leads to image stretching along $y$ (cp. Fig. 1(a) with Fig. 4(a)) [7]. The values

**R. V. Lapshin**

Table 5. Mean velocities of total drift, thermal drift, and creep. The values in parentheses are obtained by summing the *y*-components of velocities of thermal drift and creep. Velocities are given relative to the coordinate system of the direct image.

| HOPG scan | Mean total drift velocity (Å/s) | | Mean thermal drift velocity (Å/s) | | Mean creep velocity (Å/s) | |
|---|---|---|---|---|---|---|
| | $\bar{v}_x$ | $\bar{v}_y$ | $\bar{v}_x^t$ | $\bar{v}_y^t$ | $\bar{v}_x^c$ | $\bar{v}_y^c$ |
| Direct, Fig. 1(a) | 0.062 | 0.081 (0.081) | 0.059 | 0.057 | 0.003 | 0.024 |
| Counter, Fig. 1(b) | 0.069 | 0.110 (0.114) | 0.069 | 0.131 | 0.000 | -0.017 |
| Direct, Fig. 6(a) | 0.153 | 0.141 (0.142) | 0.146 | 0.103 | 0.007 | 0.039 |
| Counter, Fig. 6(b) | 0.098 | 0.138 (0.137) | 0.094 | 0.153 | 0.004 | -0.016 |

belonging to the regression surface $\bar{K}_y^r$ of the counter image (see Fig. 3(e), Fig. 5(b)), on the contrary, exceed the specified coefficient $\Delta_y$. In this case, the actual microscope step is greater than the specified one, thermal drift $\bar{v}_y^t$ is opposite to the movement, and the image is undergoing shrinkage along *y* (cp. Fig. 1(b) with Fig. 4(b)) [7]. In Fig. 5, 1st order regression curves are used since, as it was determined earlier, the 1st order is quite sufficient for the observed distortions to be precisely described.

The error caused by thermal drift and creep along the slow scan direction (see Fig. 5(b)) can be found as a difference between the known coefficient $\Delta_y$ and the regression surface profile $\bar{K}_y^r(0, y)\cos[\alpha^r(0, y)]$. Assuming that creep is finished by the end of the direct scan, the mean thermal drift of the direct scan can be calculated as follows

$$\bar{v}_y^t = \frac{\{\Delta_y - \bar{K}_y^r(0, y_{max})\cos[\alpha^r(0, y_{max})]\}y_{max}}{T}. \quad (1)$$

The numerator of the formula is an area of the hatched rectangle in the figure.

Accordingly, the mean creep velocity in *y* direction can be determined by

$$\bar{v}_y^c = \frac{\sum_{y=0}^{y_{max}} \{\bar{K}_y^r(0, y_{max})\cos[\alpha^r(0, y_{max})] - \bar{K}_y^r(0, y)\cos[\alpha^r(0, y)]\}}{T(0, y_{max})}, \quad (2)$$

where $T(x, y) = \frac{\Delta_x}{v_x}(x + 2x_{max}y) + \frac{\Delta_y}{v_y}y$ is the time since the beginning of the scan till the time the probe has reached the raster point with coordinates *x*, *y*; $v_x$, $v_y$ are the movement velocities in the raster line and from line to line of the raster, respectively (as a rule, $v_x = v_y = v$). In case of the 1st order regression surfaces, the formula (2) can be simplified to

$$\bar{v}_y^c = \frac{\{\bar{K}_y^r(0, y_{max})\cos[\alpha^r(0, y_{max})] - \bar{K}_y^r(0,0)\cos[\alpha^r(0,0)]\}y_{max}}{2T}. \quad (3)$$

The numerator of the formula is a doubled area of the hatched triangle in Fig. 5(b). The formulae for calculation of

# Drift-insensitive distributed calibration of probe microscope scanner

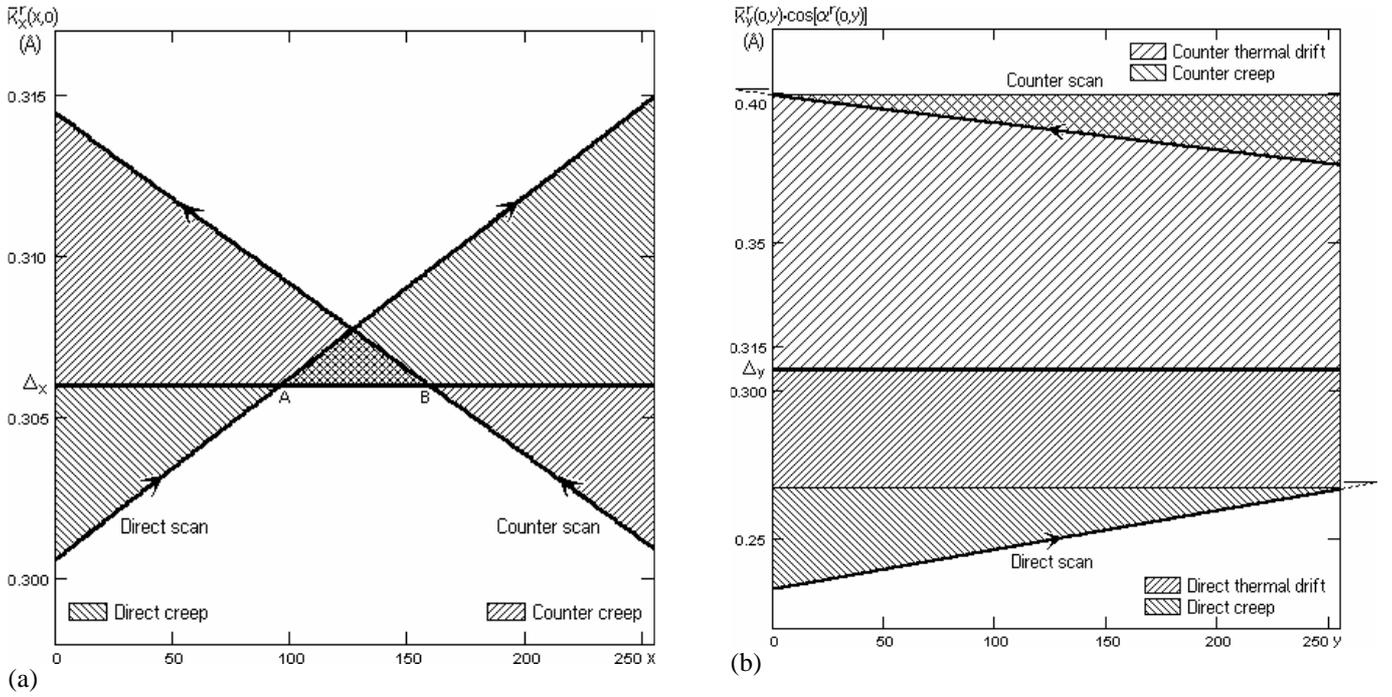

(a)
(b)

Fig. 5. Regression curves of the 1st order showing the change in microscope step while moving along the raster (a) first line, (b) first column. Such curves allow extracting information about mean velocities of thermal drift and creep.

the mean thermal drift and the mean creep of the counter scan are built the same way; the found values are given in Table 5.

Checking by formulae

$$\begin{aligned}\overline{v}_x &= \overline{v}_x^t + \overline{v}_x^c, \\ \overline{v}_y &= \overline{v}_y^t + \overline{v}_y^c\end{aligned} \quad (4)$$

shows that the obtained values (presented in parentheses in the table) are close to the total drift determined above. Apparently, the small differences are linked to the creep which does not finished by the end of the direct (counter) scan (the situation is conditionally shown in Fig. 5(b) with dotted lines) as well as with the virtual calibration which omitted peripheral areas of the scan (to avoid influence of edged distortions).

Mean thermal drift along $x$ direction is calculated according to the formula

$$\overline{v}_x^t = \frac{\sum_{y=0}^{y_{max}} \overline{K}_y^r(0,y)\sin[\alpha^r(0,y)]}{T}. \quad (5)$$

The obtained values (see Table 5) are very close to the values of the total drift. This implies that the mean creep along $x$ is very small (the trace creep is compensated by retrace creep).

Mean creep in the fast scan direction is determined as follows

$$\overline{v}_x^c = \frac{\sum_{x=0}^{x_{max}}[\Delta_x - \overline{K}_x^r(x,0)]}{T(x_{max},0)}, \quad (6)$$

where $\Delta_x$=0.306 Å is a lumped calibration coefficient along $x$ [14]. The creep values $\overline{v}_x^c$ presented in Table 5 are calculated using formula (4) since calculations using formula (6) lead to a large error.

The fact of crossing between the regression line $\overline{K}_x^r$ and the horizontal line $\Delta_x$ (point A) means that the distor-tions caused by creep in $x$ direction in this place of the scan are completely absent. One should pay attention to the significant discontinuity of the regression surfaces at the end of the direct line/scan and in the beginning of the counter line/scan that points out a large dynamic error that occurred at these raster points [10, 15].

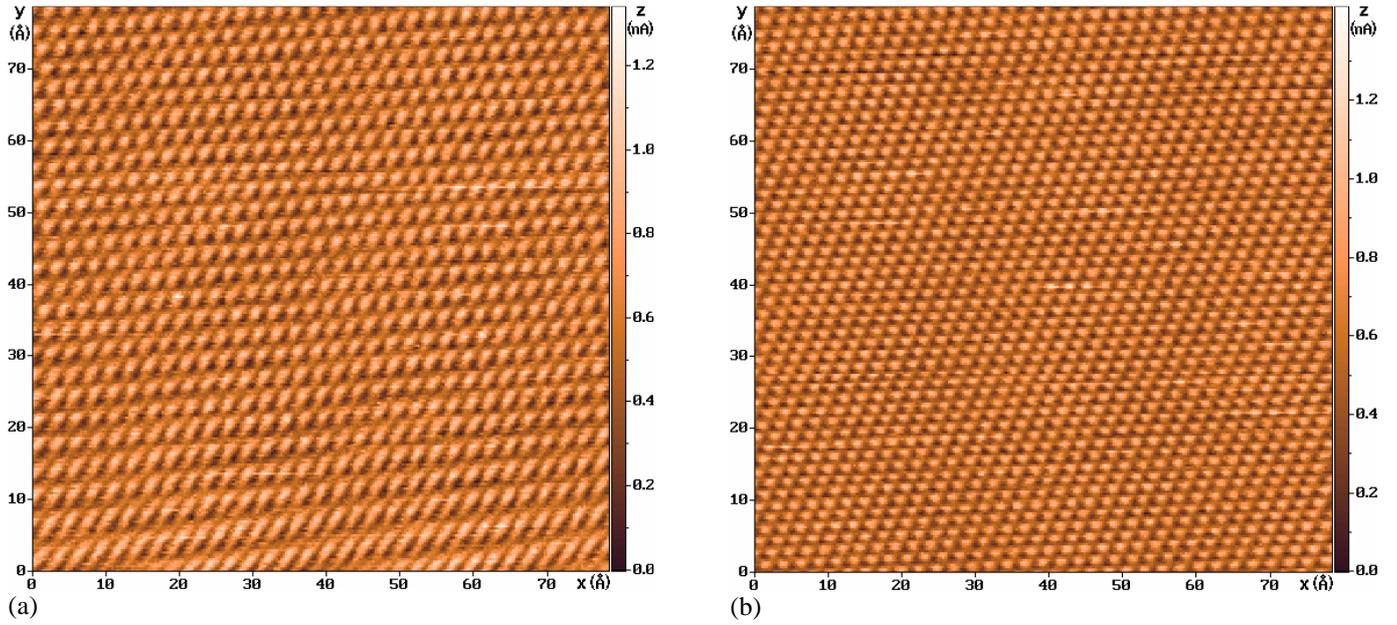

Fig. 6. Scans of atomic surface of pyrolytic graphite distorted by strong drift (a) direct image, (b) counter image. Well noticeable distortions in the beginning of the direct scan were intentionally provoked prior to the scanning by an abrupt lateral scanner offset into the initial raster point. The creep excited by the offset had stopped by the moment the counter scan began. Measurement mode: STM, constant-height, $U_{tun}$=85 mV, $I_{tun}$=750 pA. Number of points in the raster: 256×256. Scanning step size: $\Delta_x$=0.306 Å, $\Delta_y$=0.307 Å. Number of averagings at the raster point is 15. Scanning velocity $v_x=v_y$=223.1 Å/s. CSI scanning time $T_{CSI}$=6 min. Mean lattice constant equals to (a) 2.9 Å, (b) 2.2 Å which corresponds to the relative measurement error of (a) 19%, (b) 13%. The maximal value of the lattice constant at the beginning of the direct scan is almost 2.5 times greater than the nominal value.

## 4. Substantially nonlinear raster distortions

*4.1. Analysis, correction, comparison of errors of direct and counter images*

A direct scan of HOPG surface substantially distorted by drift is shown in Fig. 6(a). The strong distortion was intentionally provoked prior to the scanning by an abrupt lateral scanner offset to the initial raster point $x_0$, $y_0$ [9]. The distortion in the beginning of the scan is so large that the lattice constant along one of the close-packed directions is nearly 2.5 times greater than the nominal value.

The counter scan is given in Fig. 6(b). According to the figure, the strong creep caused by the offset has practically terminated by the moment the counter-scanning began. As it was already mentioned above, the counter scan is generally less distorted than the direct one (see Table 3) since its creep is being partially compensated by the oppositely oriented creep of the direct scan. Thus, in some cases, despite of twice increase in scanning time, the counter-scanning is to be performed instead of conventional scanning even if no following CSI correction [7] is planned.

Since all coefficients $<\overline{K}_x>$ in Table 1 are approximately equal and close to the lumped coefficient $\Delta_x$=0.306 Å in spite of different degree of the image distortions, the raster line itself is distorted just insignificantly during the raster scanning. The fact is that the movement along a line is comparatively fast, therefore the slow drift has no time to distort it notably; and so the creep induced in the line at forward trace is compensated by the creep induced at retrace. The bulk of the drift-induced distortions along $x$ direction is related to small relative shifts of

# Drift-insensitive distributed calibration of probe microscope scanner

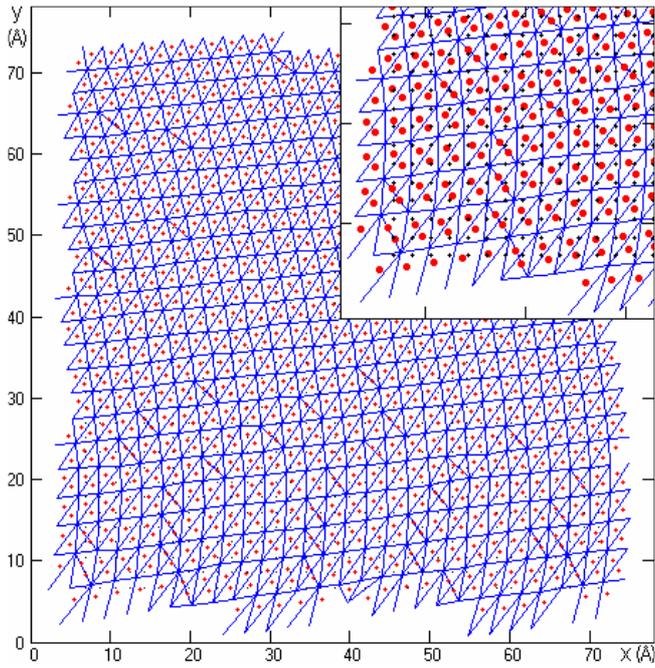

Fig. 7. LCS positions for which during the virtual distributed calibration LCCs and local obliquity angles were obtained. The initial net 36×36 nodes, step 6 positions. Carbon atoms are used as features. Number of processed apertures 1296. Number of found LCSs 1233.

lines being accumulated, which is accounted for by the obliquity angle $\alpha$. It is interesting to note that the coefficient $<\overline{K}_x>$ for such heavily creep-distorted direct image as the one shown in Fig. 6(a) has a very small distinction from the coefficients $<\overline{K}_x>$ of other much less distorted images.

From the above follows the known rule of calibration of an orthogonal scanner by a 1D diffraction grating. First, by fixing grating lines across the fast scanning direction, the grating profile should be measured by which the calibration coefficient $K_x$ should be determined. Then, after rotating the grating by 90 degrees and switching the fast scanning direction from $x$ to $y$, the grating profile should be measured again followed by determining the calibration coefficient $K_y$.

In Fig. 7, LCS positions are shown for which LCCs and local obliquity angles were obtained during the virtual calibration by the direct image of Fig. 6(a). Despite of very strong distortions in the beginning of the direct scan, the algorithm of distributed calibration recognized the graphite unit cells without errors and correctly determined LCCs within this part of the scan.

In Fig. 8, 2nd order regression surfaces built with the use of the joint CDB of the direct scan are shown. In accordance with the results obtained before, a strong curvature of the regression surface $\overline{K}_y^r$ might have been expected. However, actually the curvature of this surface turned out to be even notably less than the one of the surface (see Fig. 3(b)) obtained for the direct scan of Fig. 1, where the typical values of distortions of raster scanning are observed (to simplify the comparison, the scales in Figs. 8(a), (b) are set the same as in corresponding Figs. 3(a), (b)).

If the regression surfaces $\overline{K}_x^r$, $\overline{K}_y^r$ are weakly curved and the observed distortion degree of the direct scan in

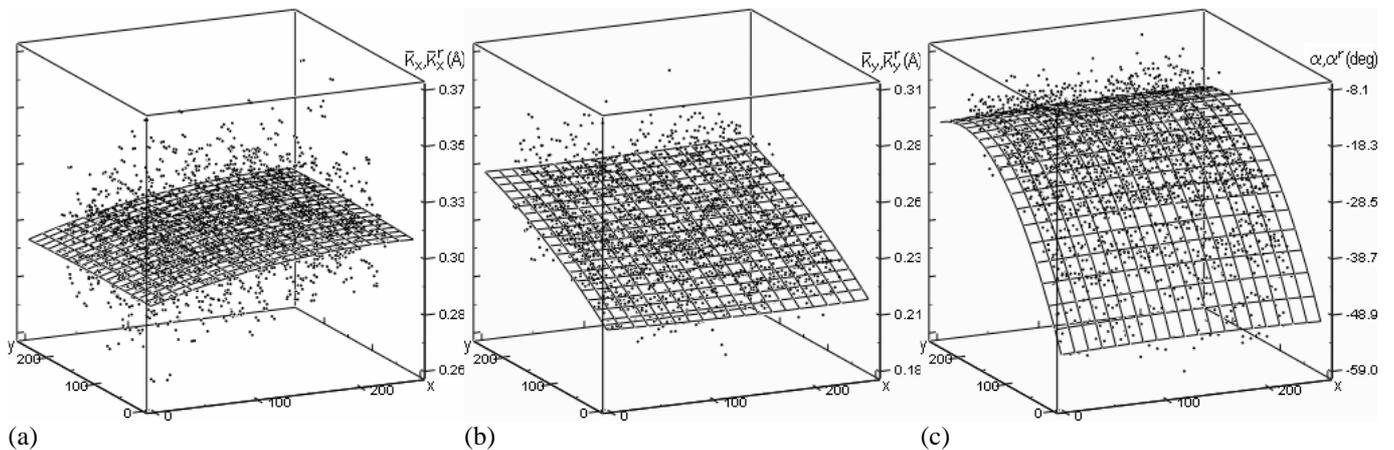

(a)      (b)      (c)

Fig. 8. Regression surfaces of 2nd order drawn through (a) LCCs $\overline{K}_x$, (b) LCCs $\overline{K}_y$, (c) local obliquity angles $\alpha$ of the direct image.

**R. V. Lapshin**

Fig. 6 is high, then the curvature of the regression surface $\alpha^r$ is expected to be strong. Indeed, in Fig. 8(c) we observe such a curvature (see also Table 3). Moreover, in comparison with the surface in Fig. 3(c), this curvature is not only significantly larger by absolute value but also has an opposite sign. Thus, in the considered distortion it is the obliquity angle that undergoes the main change, which is reflected in the image as a strong nonlinear shift of the raster lines relatively to each other.

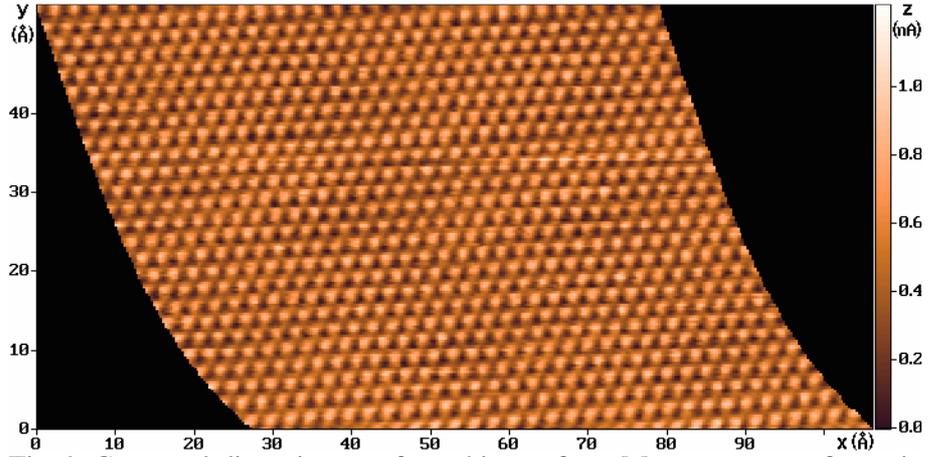

Fig. 9. Corrected direct image of graphite surface. Mean constant of atomic lattice is equal to the nominal value 2.464 Å. Number of image points 346×175. "Size" of an image point $\Delta_x=\Delta_y=0.3065$ Å.

The corrected direct scan is shown in Fig. 9. The correction of the direct scan is carried out by formulae (12) given in Ref. 4 by using regression surfaces of the 1st order for $\overline{K}_x^r$ and the 2nd order for $\overline{K}_{yx}^r$, $\overline{K}_{yy}^r$. Correction of the counter scan is carried out by formulae (12) by using regression surfaces of the 1st order for $\overline{K}_x^r$ and the 4th order for $\overline{K}_{yx}^r$, $\overline{K}_{yy}^r$. Control of the correction results by mean value of the lattice constant was conducted by means of the virtual FOS and it proved that in any part of the direct image each carbon atom has six neighboring atoms forming a regular hexagon with a side equaled to the lattice constant set for calibration (see Table 4). A somewhat higher error of lattice constant determination in the counter image could be explained by the fact that the features were represented in the counter image by a smaller number of points than in Fig. 1(b) because of a stronger squeezing of the image along *y*.

The validity of the performed nonlinear correction is confirmed by virtual distributed calibrations by the corrected images (see Table 4). Map of LCS positions, regression surfaces, and a corrected topography of the counter scan are not given because they have insignificant distinctions from the images obtained earlier for the counter scan in Fig. 1. Mean velocities of drifts and creeps during the scanning are given in Table 5.

## 5. Discussion

The virtual mode can be applied for correction of nonlinear distortions of surface topography when the surface image contains elements quite evenly distributed over the scan area and with *a priori* known sizes or distances between them [5, 16]. For example, by using the surface of a standard as a substrate, one may deposit detached objects under investigation on this surface, carry out regular scanning, perform virtual distributed calibration by the known observable elements of the standard, and then correct the obtained image according to the acquired CDB. Moreover, by performing a regular scan of a standard surface under the same conditions and with the same parameters as a particular unknown surface, it becomes possible to correct the unknown surface. The correction is carried out by the CDB acquired during virtual distributed calibration by the obtained image of the standard.

## Acknowledgments

This work was supported by the Russian Foundation for Basic Research (project 15-08-00001) and by the

# Drift-insensitive distributed calibration of probe microscope scanner

Ministry of Education and Science of Russian Federation (contracts 14.429.11.0002, 14.578.21.0059). I thank O. E. Lyapin and Assoc. Prof. S. Y. Vasiliev for their critical reading of the manuscript; Dr. A. L. Gudkov, Prof. E. A. Ilyichev, Assoc. Prof. E. A. Fetisov, and late Prof. E. A. Poltoratsky for their support and stimulation.

**R. V. Lapshin**